\documentclass{aa}  
\usepackage{graphicx}
\usepackage{txfonts}
\begin{document}
   \title{Physical parameters of pre-main sequence stars in open clusters}

   \author{Antonio J. Delgado\inst{1}, Emilio J. Alfaro
          \inst{1}
          \and
          Jo\~ao L. Yun\inst{2}
          }

   \offprints{A.J. Delgado}

   \institute{Instituto de Astrof\'\i sica de Andaluc\'\i a, CSIC, \\
    Apdo 3004, 18080-Granada, Spain\\
              \email{delgado@iaa.es},\email{emilio@iaa.es}
         \and
Centro de Astronomia e Astrof\'{\i}sica da Universidade de Lisboa,
Observat\'orio Astron\'omico de Lisboa, Tapada da Ajuda,
1349-018 Lisboa, Portugal \\
             \email{joao.yun@oal.ul.pt}
             }

   \date{Received ; accepted}

 \abstract
{The search and analysis of pre-main sequence (PMS) stars in Galactic young open clusters (YOCs)}
{The aims are twofold: To determine the ages, masses and spatial distribution of PMS members in young open clusters, and to check and compare the performances of different model isochrones}
{We compare $UBVRI$ photometric observations to theoretical isochrones in the photometric diagrams. The comparison simultaneously provides membership assignments for both main sequence (MS) and PMS stars, and estimates for the physical properties of the cluster candidate members: masses, ages, and spatial distribution inside the cluster}
{The photometric measurement of an average $(U-V)$ excess, considered as an indication of the presence of accretion disks, is considered prior to membership assignment. This photometric excess is correlated with cluster age, suggesting a vanishing of disks, able to show up in $(U-V)$ excess, at ages around 5 Myr. We find more satisfactory values of measured masses from comparison to models in the colour-magnitude (CM) diagram than in the theoretical HR plane. The obtained cluster mass functions show a marginal steepening with cluster age. Significant variations in the mass function slopes are found with the models used in member selection. The clusters \object{NGC\,3293} and \object{NGC\,2362} are found to have mass functions flatter than the Salpeter slope for all models considered. The relation between the calculated dispersion of PMS age and the characteristic clustering scale of the cluster shows an interesting agreement with previous findings in star forming regions in a wide range of scales. Finally, the ratio of characteristic clustering scales for PMS candidate members in different mass ranges can be interpreted as suggesting mass segregation, in the sense of a relatively wider spatial distribution for the lower mass members in older clusters}
{The relations between the different cluster parameters show that the procedure applied to assign cluster membership, and to measure physical parameters for the selected members, is well founded. } 



   \titlerunning{Physical parameters of PMS stars in YOCs}
   \authorrunning{A.J. Delgado et al.}

   \maketitle
%

\section{Introduction}

The primary aim of a general investigation of star formation is to obtain reliable measurements of the distribution of the forming objects in mass, age and spatial structure. A simultaneous aim of the investigation is to constrain the models used in the determination of these physical parameters.

Further objectives may include details of the time evolution of these properties, and their dependence on other parameters, such as the chemical composition, the total mass of the initial cloud, or the environmental conditions. But the fundamental aim is to tune the theoretical models, and to obtain accurate values of masses and ages.

In this context, stellar clusters are the most reliable tool to measure distances, masses and ages of stars. Young open clusters (YOCs), in particular, serve to increase our understanding of stars in early evolutionary phases, including the pre-main sequence (PMS) evolution. Our work deals with YOCs of ages around 10 Myr, which offer some clear advantages to the study of the star formation process in clusters. These objects are usually not embedded in the remnants of the dust and gas clouds where they formed, and can be studied with multiwavelength photometric observations, covering the optical and the infrared range. For these clusters, reliable determinations of distance and absorption are possible, which greatly help in further determinations of physical parameters of the PMS cluster members, such as mass and age.

In recent years, detailed new observations of some particular YOCs have enlarged the data available to test and constrain model predictions. Primarily, X-ray detections and H$\alpha$ emission, together with spectroscopically determined spectral types, provide assessment of cluster members, and, subsequently, models are used to obtain masses and ages (Flaccomio et al. 2006, F06 in the following; Dahm et al. 2007, D07 in the following). In this context, several sets of PMS isochrone models have been published, which cover different ranges in star mass and other physical parameters (see Hillenbrand \& White 2004, H04 in the following). 

Recent studies have compared models to observations on the basis of synthesized clusters, simulated from different models and with assumed contributions from the expected sources of uncertainty (Hillenbrand et al. 2008). Another approach to assess age determinations of theoretical models has been advanced recently, based on the analysis of stellar pulsation of PMS stars and comparison with predictions of model interiors (Zwintz et al. 2008). 

The measurement of physical parameters such as mass and age is achieved through the comparison of observations with evolutionary stellar interior models in the HR diagram. This comparison between models and observations can be performed in two ways. The first consists of calculating luminosity and effective temperature from observed colours, and afterward the physical parameters are read from the theoretical isochrones in the HR diagram (examples for \object{NGC\,2264} by Rebull et al. 2002, R02 in the following; F06). This approach is used in the methods for age measurement reviewed by Naylor et al. (2009). This procedure has the advantage of a more accurate consideration of extinction for the individual stars. But the comparison to models needs additional information, such as spectral types and, most important, membership confirmation, which is usually lacking. The second approach consists of translating the theoretical isochrones to colours and absolute visual magnitudes, and of comparing to observations in the photometric diagrams (D07, also for \object{NGC\,2264}). For a general sample of clusters, the conversion from theoretical luminosity and effective temperature to photometric colours is preferred, in particular if the extinction does not show too high values or degree of variability. The transformed isochrones can then be used as reference lines to measure colour excess and distance. Interestingly, this approach allows the simultaneous study of three issues, a) assignment of cluster membership, b) determination of the physical parameters for the candidate members, and c) test of the performances of different evolutionary calculations. 

In this paper we present the results of the methodology used to obtain basic physical information on the PMS member stars in YOCs. It follows the second approach described in the previous paragraph. The clusters are selected to be in an estimated age range between 5 and 30 Myr, located in a narrow range of Galactic longitude, little affected by reddening, and located at relatively close Sun distances. With these criteria we expect to minimize the influence of factors which introduce uncertainty in the determination of cluster parameters. The selected clusters can be observed well with small to medium size telescopes, mostly in a single campaign, to make the photometry as homogeneous as possible. With the mentioned requirements, we expect the detection of PMS cluster members down to at least early K spectral types.

The use of $UBVRI$ photometry, adequately compared to theoretical isochrones, provides homogeneous samples of candidate members, giving at the same time a measurement of their ages and masses. This can be used to investigate basic features of the star forming process, as described by the spatial distributions of the PMS members inside the cluster, the mass function and the structure in age. Beyond the obtaining of specific relations between parameters, our main aim is to show the ability of our procedure to advance in the desired direction: Defining PMS member samples in young clusters all accross the Galactic disk, and opening the possibility of an improved and systematic checking of PMS evolutionary models. 

We show how the use of $UBVRI$ photometry only, adequately compared to theoretical isochrones, is capable of providing useful results on the basic physical parameters of PMS stars. Our method provides homogeneous samples of candidate members, giving at the same time a measurement of their ages and masses. This is used to investigate basic features of the star forming process, as described by the spatial distributions of the PMS members inside the cluster, the mass function and the structure in age.

In Sect.~2 we review the procedure applied to estimate PMS membership through fitting to model isochrones, and describe the comparisons between the members selected using our method and published results on some YOCs. In Sect.~3, the results of our method, applied to 11 southern YOCs in the age range 5 to 30 Myr are presented. In Sect.~4, the main results and conclusions are outlined. 

\section{Membership determination}

In the present study we focus on the results for 11 southern young open clusters (YOCs), included in our previous papers (Delgado et al. 2006, Delgado, Alfaro \& Yun 2007. DAY07 in the following). 
The clusters selected are listed in Table~\ref{t1}, together with their Galactic coordinates ($l,b$ in degrees), colour excesses, distance moduli, and ages with uncertainty and indication of the number of stars used in the age calculation.

\begin{table}
\caption{\label{t1}Clusters observed.}
\centering
\begin{tabular}{lrrcccr}
\hline\hline
 Cluster & $l$ & $b$ & $E(B-V)$ & $DM$ & $logAge$ & $N$  \\
           & deg & deg &          &      &    yr    &    \\
\hline
 \object{NGC\,2362}      & 235.65 & $-$3.84 & 0.12 & 10.8 & 6.5 $\pm$ 0.08 & 6  \\
 \object{NGC\,2367}      & 238.18 & $-$5.55 & 0.38 & 12.3 & 6.6 $\pm$ 0.13 & 5  \\
 \object{NGC\,3293}      & 285.85 &  0.07 & 0.29 & 12.0 & 6.8 $\pm$ 0.07 & 31 \\
 \object{Collinder\,228} & 287.52 & $-$1.03 & 0.32 & 12.0 & 6.7 $\pm$ 0.10 & 4  \\
 \object{Hogg\,10}       & 290.80 &  0.10 & 0.36 & 11.9 & 7.0 $\pm$ 0.18 & 5  \\
 \object{Hogg\,11}       & 290.89 &  0.14 & 0.30 & 11.9 & 6.8 $\pm$ 0.02 & 4  \\
 \object{Trumpler\,18}   & 290.99 & $-$0.14 & 0.30 & 10.6 & 7.5 $\pm$ 0.29 & 3  \\
 \object{NGC\,3590}      & 291.21 & $-$0.18 & 0.52 & 11.7 & 7.2 $\pm$ 0.07 & 9  \\
 \object{NGC\,4103}      & 297.57 &  1.18 & 0.32 & 11.5 & 7.3 $\pm$ 0.06 & 11 \\
 \object{NGC\,4463}      & 300.65 & $-$2.01 & 0.39 & 11.1 & 7.3 $\pm$ 0.22 & 5  \\
 \object{NGC\,5606}      & 314.84 &  0.99 & 0.49 & 11.8 & 7.0 $\pm$ 0.11 & 7  \\
\hline
\end{tabular}
\tablefoot{
Column 1: Cluster name. 2: Galactic longitude. 3: Galactic latitude. 4: Colour excess. 5: Distance modulus. 6: Logarithmus of age (yr). 7: Number of stars used in the age calculation
}
\end{table}

The comparison with the zero-age main sequence (ZAMS) and to Padova isochrones by Girardi et al. (2002) in the colour-magnitude (CM) diagrams provides mass estimates for the MS cluster members. The evolved MS members (designed in the following as postMS) are used to calculate the cluster ages and uncertainties listed in Table~\ref{t1} (DAY07). Values for the distance, colour excess and age of \object{Hogg\,10} and \object{Hogg\,11} have been recalculated in this work. The uncertainty of the age values, given by the standard error of the mean in each case, has typical values between 0.1 and 0.3 in the logarithm, but the age values themselves depend on the model isochrones used. 

\subsection{UV excess of PMS cluster members}

Prior to determining the membership, we analyze the possible presence of ultraviolet excess in the candidate members. In R02 it is suggested that several stars among the PMS cluster members in \object{NGC\,2264} exhibit excess continuum emission in ultraviolet wavelenghts, which should be originating in accretion disks. This is actually one of the disk signatures they explore, together with near infrared (NIR) excess and H$\alpha$ emission. They claim the excess to be present in the $U$ band, and absent from $R$ and $I$ bands. With their own spectral classification, they use the spectral-type related intrinsic colours to estimate the amount of ultraviolet excess, as observable in the $(U-V)$ colour index when plotted versus $(R-I)$ in the corresponding colour-colour (CC) diagram.  The so called $UV$ excess is defined as the difference between the dereddened $(U-V)$ and the intrinsic $(U-V)$, obtained from the spectroscopically derived spectral type.

%
   \begin{figure}
   \centering
   \includegraphics[angle=0,width=10cm]{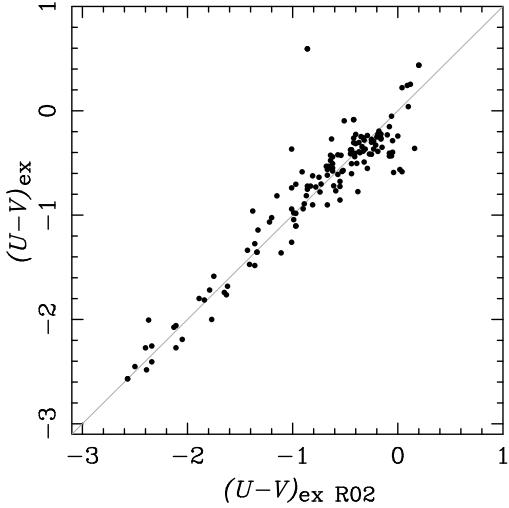}
      \caption{Excesses in $(U-V)$ for PMS members in \object{NGC\,2264} calculated from photometry (see text), plotted versus spectroscopic $(U-V)$ excess by R02. The continuous line shows the slope 1.}
         \label{UVcalc}
   \end{figure}
%

We reproduce these results on the basis of photometry, an approach which was discussed by Rebull et al. (2000) and R02. It is based on the assumption that the presence of an accretion disk does not produce any excess in the $R$ and $I$ bands, so that the red colour indices $(R-I)$ and even $(V-I)$ are free of this effect. Here we assume the median colour excess $E(B-V)$ for calculated MS members (see DAY07), together with the extinction law by Fitzpatrick (1999) to provide a colour excess $E(R-I)$ and the corresponding intrinsic colour $(R-I)_0$. This intrinsic colour is calculated for every star, and provides a spectral type estimator, which can be used in the same way as in R02 to calculate the $(U-V)$ excess.

In Fig.~\ref{UVcalc} we show the results of this calculation for cluster PMS members, plotted versus the values listed by R02. The plot shows a good performance of the assumption, which allows us to estimate $UV$ excesses in this way for any observed cluster, for which only photometric measurements are obtained, and membership and spectral type information is lacking. Excesses in $(U-B)$ and $(B-V)$ are calculated with the relations $(U-B)_\mathrm{ex}=0.76\times(U-V)_\mathrm{ex}$, $(B-V)_\mathrm{ex}=0.24\times(U-V)_\mathrm{ex}$, obtained from linear correlations to the calculated $(U-V)$ excesses in \object{NGC\,2264}. We follow the considerations by Rebull et al. (2000), and apply only those excesses which correspond to $(U-V)_\mathrm{ex}<-0.4$.

\begin{figure}
   \centering
   \includegraphics[angle=0,width=10cm]{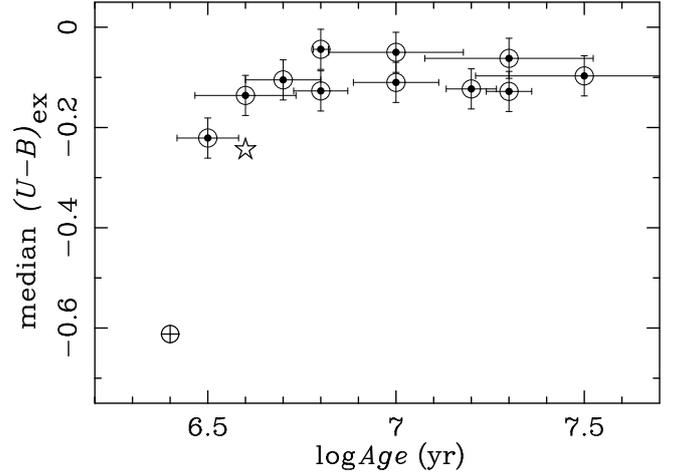}
      \caption{Median $(U-B)$ excess for the stars with negative calculated value (see text), plotted versus cluster age, obtained from quantitative comparison to Padova isochrones. The plotted error bars reproduce the standard error of the mean in each case. Values for \object{NGC\,2264} (crossed circle), and \object{NGC\,1893} (star) are also plotted for comparison
}
         \label{UVage}
   \end{figure}

The influence of the $UV$ excess is only appreciable for the youngest clusters in our sample. In Fig.~\ref{UVage} we show a plot of the median $UV$ excesses versus cluster age. We include values for the clusters \object{NGC\,2264} (2.5  Myr given by D07) and \object{NGC\,1893} (4 Myr, adopted by Sharma et al. 2007). The error bar in the $UV$ excess reproduces the standard error of the mean of the differences between photometric and spectroscopic values in \object{NGC\,2264}. 

%
%

A trend with age can be surmised in Fig.~\ref{UVage}, which suggests an age limit of around 5 Myr for the presence of disks, capable of producing $UV$ excess. We recall in this context results obtained previously, with time scales for disk dissipation from 4 to 6 Myr (Haisch et al. 2001; Sicilia-Aguilar et al. 2005).

\subsection{Fitting to model isochrones}

The procedure to select PMS cluster members has been explained in detail in DAY07, and we summarize it here. The measurement of colour excess and distance for main sequence (MS) cluster members is carried out by fitting to the ZAMS. These values of reddening and distance, estimated from all selected MS candidate members, are then used as reference values to assign membership to all measured stars. The theoretical PMS isochrones converted to the photometric diagrams are used in this assignment as reference lines to measure colour excess and distance. Membership is established when the measured colour excesses and distance coincide, within the errors, with the reference values and any particular star can be assigned as PMS member from the comparison to PMS isochrones in several CM diagrams. Furthermore, the isochrones which provide membership assignment also provide a mass and age value for the candidate members, calculated as averages of the values extracted from all assignments.

We introduce here some improvements in the membership determination with respect to DAY07. 

First, the estimate of reddening and membership for those stars without valid $U$ measurement takes into account the colour excess values calculated for all stars with measurement in all five $UBVRI$ colours, and with a membership assignment. Second, we consider the possible presence of ultraviolet excess in the stars' colours, as has been explained in the previous Sec. 2.1. As we have seen, this factor turns out to be of little importance for cluster ages above 6 Myr. Third, in relation to DAY07, the use of Yale PMS isochrones (Yi et al. 2001) is added. The four models considered are by D'Antona \& Mazzitelli (1997), Palla \& Stahler (1999), Siess, Dufour \& Forestini (2000) and Yi et al. (2001), referred to in the following as D97, P99, S00 and Y01, respectively. All of them are converted to the CM diagrams with the relations from Kenyon \& Hartmann (1995).

A catalogue with the photometric results for the 27040 stars in the 11 cluster fields  is contained in a catalogue available at CDS and from {\tt http://ssg.iaa.es}. It supersedes the members list published in DAY07. The catalogue contains the following information. The cluster identification in the OCL catalogue in column 1. The 2MASS identification (2) and our own identification (3); Right ascension (4) and Declination (5) for the Epoch 2000. Colour indices with errors, $V, \sigma(V), (U-B), \sigma(U-B), (B-V), \sigma(B-V), (V-R), \sigma(V-R), (V-I), \sigma(V-I)$ (6 to 15). Type of membership, if any, and the values of Age, mass, luminosity and effective temperature calculated with the PMS models of D97 (16-20). The same as columns 16-20, from the PMS models of P99 (21-25). The same as columns 16-20, from the PMS models of S00 (26-30). The same as columns 16-20, from the PMS models of Y01 (31-35). 

The procedure for membership assignment has in principle two main shortcomings. First, the required conversion from luminosity and effective temperature to intrinsic colours and bolometric corrections is not necessarily the same as the one used for MS stars. This difficulty is also present when converting observed colours to theoretical quantities. The differences are, however, of remarkable importance for PMS stars of class II and younger (Delgado et al. 1998). For class III objects the assumption of  validity of the calibration used for MS stars is a reasonable approximation (Kenyon and Hartmann, 1995). 

Second, the estimate of absorption for every individual star is necessary to calculate distance. For many stars in the cluster field, and in particular the faintest ones, this estimate is probably affected by undetected variations, and might lead to inaccurate estimates both for the membership, and for the mass and age of the candidate members. It is important to remark that this source of uncertainty is unavoidable for clusters where information about spectral type is absent. This is indeed the general case. Follow-up spectroscopic observations are therefore of importance, both for assessing the membership estimates and simultaneously determining the actual differences between the performances of the different models. 

Two more remarks need to be made here. First, in some clusters the procedure to select MS members (DAY07) assigns MS membership to an appreciable number of stars which are fainter than the brightest PMS selected members. These have been discussed by Delgado et al. (2010) in the case of another cluster of interest, Dolidze 25, located inside a larger star-forming region Sh~2-284. In this context we mention the recent work by Beccari et al. (2010), where progressive star formation in the young galactic super star cluster \object{NGC\,3603} is proposed, with 1/3 of PMS members found to be older than 10 Myr, while the ages of the remaining members range from 1 to 10 Myr. For the clusters in our sample, further spectroscopic assessment is necessary to confirm that these stars are actually cluster members. Their MS membership is included in the final catalogue, but they are not considered in the analysis of the results in the present paper. Second, as we have mentioned, the PMS membership assignment can be obtained with respect to isochrones in 2, 3 or 4 CM diagrams. Unless otherwise stated, in the following discussions we will use the term "PMS member" to name a star with uncertainties in all photometric indices smaller than 0.05, and with membership assignment in at least 3 CM diagrams (3CM-PMS membership assignment).

\subsection{Comparison to published results}

Previous studies contain lists of X-Ray sources in the cluster \object{NGC\,2362}, (Delgado et al. 2006, Damiani et al. 2006), and stars with H$\alpha$ emission or Li~6707\AA ~absorption (Dahm 2005). Stars with any of these features and located in the area covered by the isochrones in the CM diagrams are considered cluster members. We name them P-members. 85$\%$ of these P-members are selected in our procedure, using the S00 models. On the other hand, 70$\%$ of our PMS members with S00 models are also P-members. This gives a coincidence (agreement of assignments in both directions) of 60$\%$. This percentage decreases to around 40$\%$ with P99 and Y01 models.

In this context, a relatively higher degree of agreement is found for PMS candidates of later spectral types. A similar result is found for \object{NGC\,1893}, another young cluster located at a relatively farther distance in the direction of the Galactic anticenter. We have also tested our method for this cluster  using the $UBVRI$ photometry by Sharma et al (2007). We consider as PMS P-members all stars in the Sharma et al. photometry which are also classified as class II and III sources by Caramazza et al. (2008). Our method of membership assignment with S00 models assigns PMS membership for 74$\%$ of these P-members. This percentage turns out to be higher for stars fainter than $V=17$ (76$\%$) than for stars brighter than this (71$\%$). This suggests that PMS stars of earlier types, around AF, might be more difficult to detect in X rays as compared to PMS stars of GK types (Damiani et al. 2006).

Finally, we used the published data on \object{NGC\,2264} as a template to analyze the outcome of our procedure. Its age is close to 3 Myr (Sung 2004, D07) and it is known to host a well-populated sequence of PMS stars, ranging from spectral types AF down to the less massive T-Tauri stars. Since the first study by Walker (1956), numerous works have been devoted to this cluster (Dahm 2008 contains a complete list of references). $UBVRI$ photometry has been published by several authors. We use here the results by R02 and the UBV photometry by Walker (1956) for stars with $V\leq14$. The results on PMS membership, mass and age by R02, and also by F06 and D07 are considered. The cross-identifications between different studies and the basic photometric information are taken from WEBDA ({\tt http://www.univie.ac.at/webda/}). For the distance of \object{NGC\,2264} we adopt the value given by Baxter et al. (2009)

The application of our procedure to the combined photometric results from Walker (1956) and R02 provides 2CM-PMS membership assignment for the 91$\%$ of the joint member samples by R02, F06 and D07. The percentage decreases to 63$\%$ when considering 3CM-PMS membership. The reason for this is probably the highly variable absorption in the cluster from MS to PMS members, which makes the colour excess estimate for PMS candidates on the sole basis of photometry uncertain. In this comparison we used S00 isochrones for metallicity $Z=0.02$, which are those used by R02 to obtain their mass values. 

The physical parameters of the PMS member samples in \object{NGC\,1893} and \object{NGC\,2264} mentioned above, as well as the MS members obtained from the same photometric studies cited (Sharma et al. 2007; Walker 1956; R02), will be used as examples in the rest of the paper. 

\section{Physical parameters of PMS candidate members}

We now turn to a discussion of the results obtained for the physical parameters of the assigned member samples. The general results for the clusters are listed in Table~\ref{t3}. All the results in this table for which the use of models is involved are obtained with S00 models.

\begin{table*}
\caption{\label{t3}Physical parameters of the observed clusters.}
\centering
\begin{tabular}{lrccccccccc}
\hline\hline
Cluster        & $(U-B)_\mathrm{ex}$ & $N_\mathrm{MS}$ & $N_\mathrm{PMS}$ &        MF      &    MF$_\mathrm{PMS}$    & Age$_\mathrm{PMS}$  & $\Sigma$  & $\Delta$   & $\Delta_\mathrm{LowM}$  & $\Delta_\mathrm{HighM}$  \\
               &            &        &         &                &                &     Myr    &     Myr   &     pc   &        pc       &         pc       \\
\hline
\object{NGC\,2362}      &  $-$0.221  &  28  &   89  & $-$0.74 $\pm$ 0.21 & $-$2.30 $\pm$ 0.31 &    5.44    &    1.95   &    1.82  &      1.78       &       1.86       \\
\object{NGC\,2367}      &  $-$0.136  &  11  &  132  & $-$1.13 $\pm$ 0.29 & $-$1.95 $\pm$ 1.12 &    5.48    &    2.57   &    4.26  &      4.37       &       4.07       \\
\object{NGC\,3293}      &  $-$0.127  &  88  &  231  & $-$0.64 $\pm$ 0.19 & $-$1.63 $\pm$ 0.42 &    6.36    &    2.45   &    3.53  &      3.77       &       3.32       \\
\object{Collinder\,228} &     0.105  &  16  &  305  & $-$1.93 $\pm$ 0.25 & $-$3.21 $\pm$ 0.27 &    6.46    &    2.53   &    3.60  &      3.52       &       3.68       \\
\object{Hogg\,10}       &  $-$0.050  &  20  &  270  & $-$1.33 $\pm$ 0.47 & $-$3.35 $\pm$ 0.51 &    6.31    &    2.47   &    3.25  &      3.26       &       3.08       \\
\object{Hogg\,11}       &     0.044  &  13  &  393  & $-$1.20 $\pm$ 0.43 & $-$3.88 $\pm$ 0.72 &    6.41    &    2.42   &    3.96  &      3.93       &       3.90       \\
\object{Trumpler\,18}   &  $-$0.097  &  16  &  126  & $-$1.07 $\pm$ 0.29 & $-$2.09 $\pm$ 0.58 &    4.59    &    2.84   &    2.01  &      2.04       &       2.18       \\
\object{NGC\,3590}      &  $-$0.123  &  37  &  406  & $-$1.51 $\pm$ 0.34 & $-$2.29 $\pm$ 0.35 &    6.28    &    2.43   &    2.98  &      2.98       &       2.94       \\
\object{NGC\,4103}      &     0.128  &  49  &  214  & $-$1.31 $\pm$ 0.27 & $-$2.41 $\pm$ 0.56 &    6.18    &    2.33   &    2.92  &      2.97       &       2.78       \\
\object{NGC\,4463}      &  $-$0.062  &  21  &  151  & $-$1.35 $\pm$ 0.25 & $-$3.75 $\pm$ 0.68 &    6.23    &    2.64   &    2.43  &      2.49       &       2.23       \\
\object{NGC\,5606}      &  $-$0.110  &  21  &  215  & $-$1.53 $\pm$ 0.31 & $-$3.40 $\pm$ 0.63 &    6.03    &    2.77   &    3.27  &      3.21       &       3.32       \\
\hline
\end{tabular}
\tablefoot{Column 1: Cluster name. 2: $(U-B)$ excess defined in Sect. 2.1. 3: Number of MS+postMS members. 4: Number of PMS members, as defined in Sect.2.2. 5: Slope of the mass functions, $\delta logN_{M}/\delta logM$, for all the members of mass above the approximate completeness limit defined in Sect.3.1.2. 6: The same slope for the mass function calculated with only the PMS members. 7,8: Average age of the PMS members (7), and its rms deviation (8). 9: Average value of the distances between all pairs of PMS members. 10,11: This same quantity calculated for the distribution of PMS members of low mass (1-2 $M_\odot$) (10) and high mass (2-3.5 $M_\odot$) (11)
}
\end{table*}

\subsection{Masses of PMS members}

As explained before (see DAY07, and Sect. 2 above), the membership assignment to a particular star can happen with respect to several isochrones. This provides several age and mass values for the particular candidate, and the averages and rms deviations of all these values are taken as the mass and age values, with uncertainties, of the particular PMS candidate member. How do these values compare to previous results?. In this section we refer to previous results on PMS masses, and discuss the differences between the values obtained using various evolutionary models.

\subsubsection{Masses of PMS members in \object{NGC\,2264}}

To illustrate the results of our method to determine membership and the physical parameters of members, we discuss in this section the results on the masses of PMS members in \object{NGC\,2264}. For this cluster the comparison to isochrones can be performed on the basis of a previously established member population by independent means, and it therefore offers a good possibility to test the reliability of our procedure. 

We extract two sets of mass values from each of the published works (R02, F06, D07). First, we have the three sets of published values, all of them obtained with S00 models, which we call P-values in the following. These three sets of P-values differ from each other. Values from F06 and D07 do not differ systematically, although the dispersion of the difference is high ($M_\mathrm{D07}-M_\mathrm{F06}=0.005\pm0.26$). The R02 values are systematically lower than both the F06 and D07 values ($M_\mathrm{R02}-M_\mathrm{F06}=-0.14\pm0.26$, $M_\mathrm{R02}-M_\mathrm{D07}=-0.22\pm0.28$). A second set of mass values is obtained by shifting the transformed model isochrones in the CM $V,(B-V)$ diagram, according to the published values of distance and absorption for every individual star, and reading its mass from the isochrone at nearest distance. The mass values from this second procedure are named T-values. P- and T-values should in principle be similar to each other, since they are obtained from the same observed colours, and are compared to the same isochrone models. However, they differ significantly. 

%
   \begin{figure}
   \centering
   \includegraphics[angle=0,width=10cm]{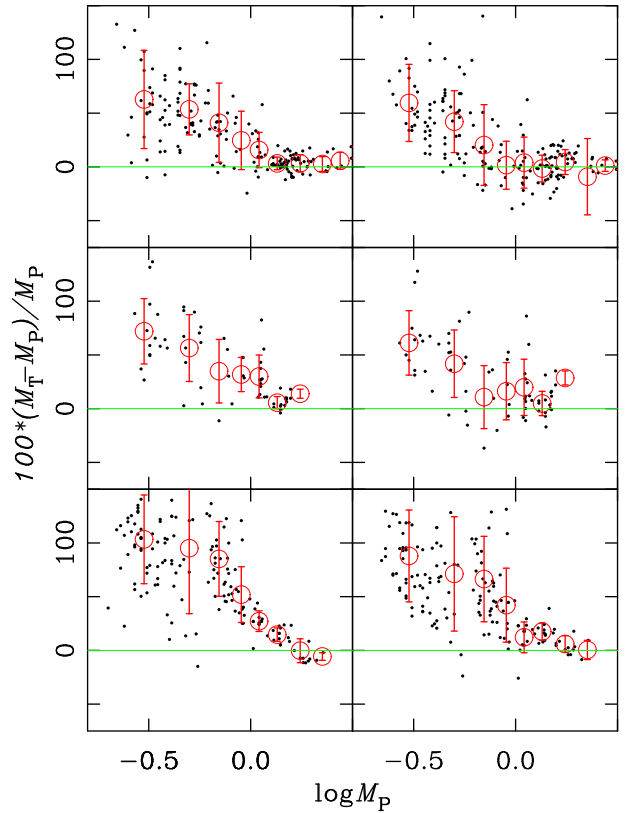}
   \caption{Percentage difference between the masses for PMS cluster members in \object{NGC\,2264}, calculated by our procedure (T-values; see text) and those by D07, F06 and R02 (P-values), from upper to lower panels, respectively. The quantity $100\times(M_\mathrm{T}-M_\mathrm{P})/M_\mathrm{P}$ is plotted versus $M_\mathrm{P}$, in the three published cases. The panels to the left show the masses calculated with respect to S00 isochrones. The panels to the right show masses calculated with respect to Y01 isochrones. Red circles and error bars show averages and rms deviations in the corresponding mass bins. The figure is to be compared to Figs. 4 and 5 in H04}
              \label{MassPC}
    \end{figure}

The results attained by H04 on star masses from different evolutionary models show that the masses of PMS stars predicted by all models are systematically lower than dynamical masses. In Fig.~\ref{MassPC} we reproduce the comparative plots produced by H04 in their Figs. 4 and 5. Here we plot the difference of T- minus P- values, versus P- values. The plot shows that T-values are systematically higher than the P-values. On the other hand, the results by H04 show that the values from the +models (associated to our P-values here) are smaller than "correct" dynamical masses. The differences shown in our plots and in those by H04 are furthermore of the same order. This means that the T-procedure produces higher mass values than the published P- values, and therefore mass results in better agreement with the dynamical values.

This result is remarkable indeed, since the models used in the calculation of both T- and P-values are again the same (S00 for the differences shown in the left panel of Fig.~\ref{MassPC}), and the photometric colours, distance and absorption for every member are also the same in both calculations. The difference for the D07 values is especially striking, since the mass calculation in D07 follows a similar procedure to our T-procedure. 

The differences shown in Fig.~\ref{MassPC} suggest that the mass results derived from a comparison of observations to models in the CM diagram offer better performances than the comparison in the theoretical luminosity versus effective temperature HR diagram, although these differences may change, depending on the particular calibration and model used. This is the case for different calibrations $T_\mathrm{eff}+L$ from magnitudes (H04) and also occurs when using different models, or different formulas to translate the theoretical quantities to magnitudes and colours. This is exemplified in the right panels of Fig.~\ref{MassPC}, where the T-values calculated from comparison to Y01 isochrones are used. In this case both isochrone sets, S00 and Y01,  have been transformed with the same calibration to the CM plane, and the differences observable between the two panels can therefore be ascribed to differences between the S00 and Y01 models.

\subsubsection{Mass functions}

We now discuss the results of the masses derived in our work, which can be associated with the so called T-values above. The clusters are at different distances, and their photometry, for the PMS candidate members in particular, has different levels of completeness. For the present purpose, we estimate this level to be at the star whose absolute magnitude in the 5Myr isochrone, added to the apparent distance modulus of the cluster ($3.1\times E(B-V)+DM$) equals the faintest magnitude of all observed stars with photometric errors in all indices below 0.05 mag. We recall that this was a condition to consider a given star as a 3CM-PMS cluster member. 

The mass functions discussed here are calculated for all member stars with a mass above the value given by this approximated completeness level. This ranges from 0.65 $M_\odot$ for \object{NGC\,2362} and \object{Trumpler\,18}, to 1.3 $M_\odot$ for \object{NGC\,5606}. The mass functions include masses estimated for MS and postMS members, obtained from the Padova isochrones (Girardi et al. 2002), also used to determine the age of the clusters (DAY07). 

\begin{figure}
   \centering
   \includegraphics[angle=0,width=11cm]{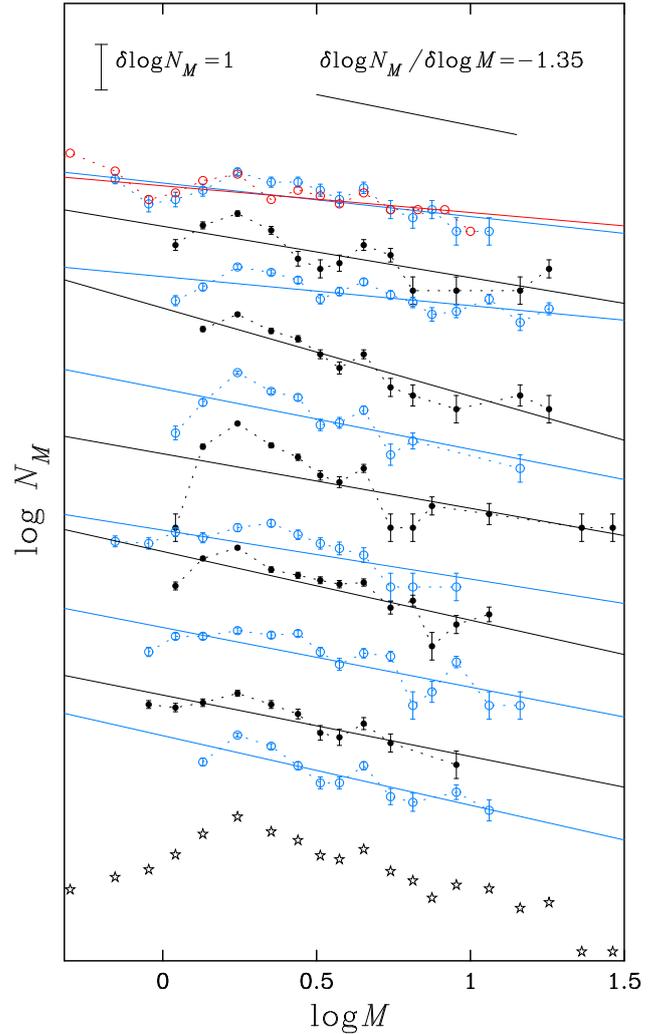}
   \caption{Mass functions for all clusters, with PMS member stars obtained from S00 models. The sequence of mass functions follows the order in Table~ \ref{t1} (ascending Galactic longitude). The size of the scale in the vertical axis, and the Salpeter slope ($-$1.35) are shown in the upper part of the plot. For \object{NGC\,2362}, the number of members with X-ray activity, H$\alpha$ emission and Li6707\AA ~absorption are plotted as red circles for comparison. The joint mass function obtained by adding the member numbers of all clusters in each mass bin is plotted at the bottom as stars 
}
              \label{MasFunTodas}
    \end{figure}

In Fig.~\ref{MasFunTodas} we plot as an example the obtained mass functions, using the S00 models. The plot includes an indication of the Salpeter slope. In all sequences plotted in Fig.~\ref{MasFunTodas} one can observe the presence of a break at around $\log M\sim0.2$ (1.5 $M_\odot$). This finding is in agreement with the results on \object{NGC\,1893} by Sharma et al (2007) and the mass function obtained by Damiani et al. (2006) from independent membership estimators, although the particular values might differ. We also note the relatively flatter slopes for the clusters \object{NGC\,2362}, also found by Damiani et al. (2006), and \object{NGC\,3293} (Slawson et al. 2007). The case of \object{NGC\,2362} is particularly interesting. It offers a good verification of our results, since we dispose of PMS member samples from different indicators, and mass functions have, moreover, been derived by other authors (Damiani 2006). In addition to our membership and mass estimates, we have calculated the mass function with the S00 masses for the PMS sample associated with X-ray sources (Delgado et al. 2006), and those with H$\alpha$ emission and Li6707\AA ~absorption (Dahm 2005). The resulting mass function is plotted in Fig.~\ref{MasFunTodas}, and shows very good agreement with the slope determined for our selected PMS sample.

%

\begin{figure}
   \centering
   \includegraphics[angle=0,width=10cm]{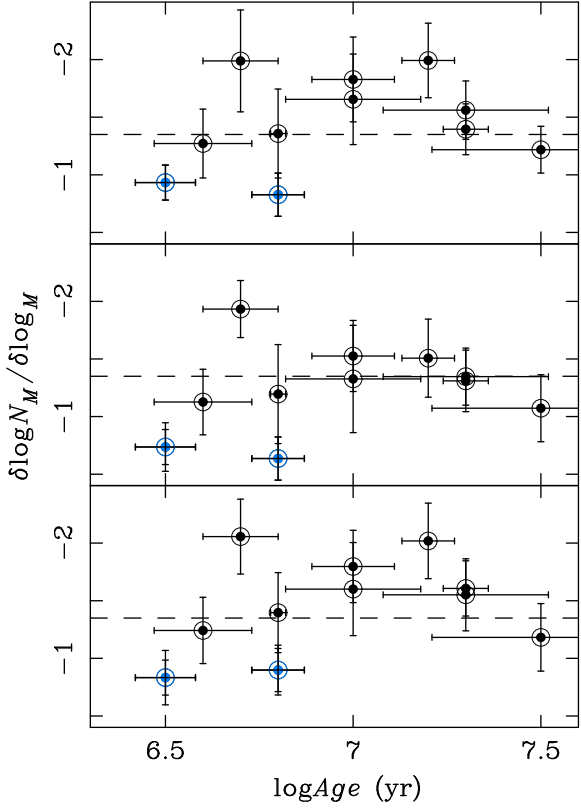}
   \caption{Slopes of the mass functions ($\delta\log N_{M}/\delta\log M$) for the 11 clusters in the sample as a function of cluster age. Members of mass above the approximate completeness limit defined in Sect.3.1.2 are considered. The slopes with PMS members using three different models are plotted. From top to down: P99, S00, Y01. Vertical error bars reproduce the slope errors from the least squares fits. Horizontal broken lines mark the Salpeter IMF slope ($-$1.35) 
}
              \label{MasFunAge}
    \end{figure}

In Fig.~\ref{MasFunAge} we plot the slopes of the mass functions $\delta\log N_{M}/\delta\log M$ versus cluster age. These are the slopes of linear least squares fits like those shown in Fig.~\ref{MasFunTodas}, obtained for the clusters in our sample with PMS members from three models (P99, S00, and Y01). The derived mass function slopes shown in Fig.~\ref{MasFunAge} show some differences between models. We observe the better agreement with the Salpeter (1955) value of the mass functions from S00 models, and a slight steepening of the slope with increasing age. In general, the differences between models need assessment with independent membership determinations. Once the refined member samples are obtained, they offer the possibility of checking the performances of the different models.

%

%
   \begin{figure}
   \centering
   \includegraphics[angle=0,width=10cm]{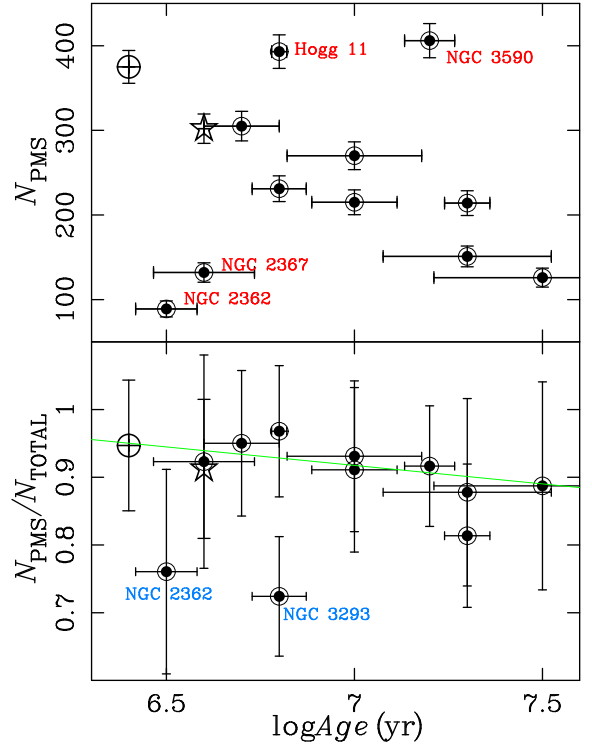}
   \caption{Number of PMS members (upper panel) and ratio of PMS to total member numbers (PMS+MS+postMS), plotted versus cluster age. The values for \object{NGC\,2264} (crossed circle) and \object{NGC\,1893} (star) are shown for comparison. The green line in the lower plot represents a linear fit to the points, excluding the clusters marked in red with their names in the upper plot. A correlation coefficient of 0.7 is obtained. Error bars in the age are listed in Table~\ref{t1}. In the vertical axis, they are calculated assuming absolute error in the numbers equal to the square root of the numbers themselves
}
              \label{NpmsvsAge}
    \end{figure}

The differences in mass functions are reflected in the relative numbers of PMS members to total number of members (PMS+MS+postMS). These numbers are listed in Table~\ref{t3}. The PMS member number, $N_\mathrm{PMS}$, and the ratio of $N_\mathrm{PMS}$ to total members' number are plotted versus cluster age in Figure~\ref{NpmsvsAge}. 

Several factors are present in the spread seen in this kind of plots. In addition to the general trend present in the sample, the uncertainties of the plotted parameters, possible biases affecting the sample and the real deviations due to peculiar properties of some objects combine to increase the spread. We consider the plot in the upper panel in Fig.~\ref{NpmsvsAge} as an example of this. The plot can be interpreted as showing a decrease of PMS member number with cluster age for the majority of clusters, with four clusters that deviate from the trend. Three of these clusters (\object{NGC\,2367}, \object{Hogg\,11}, and \object{NGC\,3590}), however, follow the slight decrease of the relative numbers which can be surmised from the plot in the lower panel. This suggests that they would be different in richness of members to the other clusters in the sample, but would exhibit a "normal" relation between PMS and total member numbers. On the other hand, the location of \object{NGC\,3293} in both plots suggests that it has a relatively larger number of MS+postMS members, in agreement with the findings by Slawson et al. (2007), whereas the location of \object{NGC\,2362} in both plots would suggest a real deficit of PMS members in the cluster, a property also detected by Damiani et al. (2006). In this context, we note that \object{NGC\,2362} is located at a relatively close distance and could not fit completely in our field of view. Some PMS members could be not included. 

Vertical error bars plotted in Figure~\ref{NpmsvsAge} are calculated with the assumption that the number of members has an absolute error given by its square root. In spite of the uncertainties, the decreasing trend with age of the relative number in the lower plot and the deviating location of the two clusters marked can be estimated. 

A final comment has to be made on the masses of the PMS candidate members. The calculation of the mass functions has also been performed for only those PMS members with masses higher than the approximated completeness limit in each cluster defined in Sect. 3.1.2 above. In these mass ranges we deal with AF spectral types for the selected PMS stars. For this samples, the slopes obtained are clearly steeper than the Salpeter value, also showing clear differences between the values obtained for masses from different isochrone models. In Table~\ref{t3} we also list the mass function slopes only for those PMS cluster members obtained with S00 models. We recall in this context the high spread in derived mass function slopes reported by Kroupa (2002), which is particularly pronounced for young clusters, and specially in the mass range 0.8 to 2.5 $M_\odot$. This overlaps with the lower half of the range of mass values for our selected PMS members.

\subsection{Age and spatial structure}

In this subsection we analyze the results of the age and spatial distributions of the PMS member sequences. As mentioned above, several factors affect the calculated parameters, with the possible undetected peculiarities of some clusters being an important source of variation. We aim to assess the reliability of the selected member samples, which should be reflected in the presence of relations between parameters with regularities clearly beyond what could be expected from an alleatory or ineffective members selection.

\subsubsection{Age and age spreads}

The measurement of star ages is always the result of comparison with evolutionary stellar models. This means that an assessment of models is required, and this can in turn only be achieved by comparison with the observations of the same stars whose age we want to measure. In other words, age measurement and model improvement have to be carried out in parallel, if not simultaneously.

Several methods have been proposed to improve our knowledge of PMS ages. The Lithium abundance and features in the CM diagram are the most recently discussed ones (Naylor et al. 2009, Cignoni et al. 2010, Cargile \& James 2010). Some of them present some unsolved problems (Naylor 2009, Jeffries et al. 2009)  and all methods need the input of an evolutionary model to go from relative to absolute age measurements. In addition to relative ages, one important issue in the age structure of YOCs concerns the actual presence of age spreads in the PMS member sequence, and how large this might be. The possible age spread is expected to reflect the presence of either instantaneous, or continuous, or episodic star formation in the cluster (Hillenbrand 2008). 

The presence of age spreads among PMS members and the age differences with the MS members in clusters is an open question. In general, the presence of age spreads in the PMS sequences is accepted, with values ranging from 2-3 Myr to more than 10 Myr (Park et al 2000; DeGioia-Eastwood et al. 2001). Some results indicate that the formation of lower mass PMS stars continues after the most massive stars have formed, although with varying efficiency, while other studies conclude that the normal sequence of formation is in fact from lower masses to higher masses (Ojha et al. 2010). Some mentions are made in this context to the possibility that the opposite case would result in inhibition of the formation of low mass stars caused by radiative feedback from the more massive members (Price \& Bate 2009; Zinnecker \& Beuther 2008). The determination of age spreads in real clusters are in any case influenced by the undetected presence of binaries, which can lead to a large overestimate of the spread (Weidner et al. 2009)

With our procedure we obtain age values with uncertainties for all assigned members, as explained in Sect. 2.2. The age of the PMS sequence in each cluster is then calculated as the average of all ages for the candidate members. The rms deviation of this mean value, $\Sigma$, is in principle indicative of the possible age spread. The age values obtained from the four models used here (D97,P99,S00 and Y01) are included in the on-line catalogue. These values are in good agreement with previous results on different YOCs (Park \& Sung 2002, Mamajek et al. 2002). These works indicate that ages from D97 are distinctly lower than those measured using the other three models, by amounts reaching several Myr. On the other hand, differences between P99, S00 and Y01 models are at most of the order of 0.5 to 1 Myr but still systematic and significant. The P99 models predict lower values than both S00 and Y01 models. These last two produce similar results. The average PMS ages of each cluster, obtained from S00 models, are listed in Table~\ref{t3}.

%
   \begin{figure}
   \centering
   \includegraphics[angle=0,width=9cm]{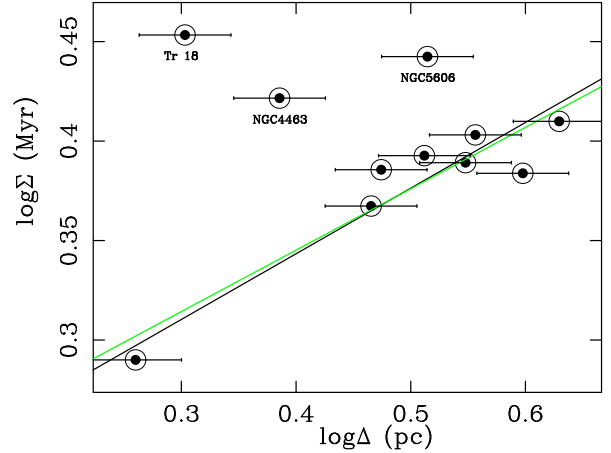}
   \caption{Relation between the rms deviation of the men PMS age ($\Sigma$) and the size of the clusters, given by the characteristic clustering scale ($\Delta$. See text) for PMS members of masses above the approximate completeness limit (see Sect. 3.1.2). Horizontal error bars are calculated with the assumption of a typical error in the distance modulus of 0.2. A bootstrapping estimate of the uncertainty in $\Sigma$ gives values similar to the size of the representative points. The straight line reproduces the slope (0.33) of the scaling relation between age spread and size proposed by Efremov \& Elmegreen (1998) for star forming regions in a wide range of sizes. The green line shows a linear least squares fit, excluding the clusters marked with their names. A slope of 0.31 and a correlation coefficient of 0.8 are obtained.}
              \label{AdvsC}
    \end{figure}

In reference to the age spreads, the use of independent membership determination from spectroscopy is especially needed to add the necessary constraints to the results from photometry and models. Without this membership confirmation, the standard deviation of the mean age, $\Sigma$, cannot be associated with a real age spread. In spite of this shortcoming, a remarkable relation is found between the calculated standard deviation and the typical clustering scale, $\Delta$, calculated as the average value of the distances between all pairs of members. This  defines the spatial concentration of the population considered (Kaas et al. 2004). In Fig.~\ref{AdvsC} $\Sigma$ is plotted versus this clustering scale, calculated in each cluster for all pairs of PMS members with mass above the approximate completeness limit defined in Sect. 3.1.2. In the figure we overplot a line of slope 0.33, of the scaling relation between age dispersion and size proposed by Efremov \& Elmegreen (1998) for clusters and associations in a wider range of scales, both below and well above the spatial scale involved in our clusters. Three clusters do not follow the relation, due probably to undetected errors in the membership assignment. For the most outstanding case, \object{Trumpler\,18}, DAY07 found signs of two clusters superimposed in the line of sight, of different ages and at well separated distances. It is one of the clusters with the largest age uncertainty in the sample. These properties probably affect the calculated spreads. Excluding the three deviating clusters, a linear fit to the data provides a slope of 0.31, with a correlation coefficient of 0.8. We consider that the agreement observed in Fig.~\ref{AdvsC} provides additional support to the ability of our procedure to detect PMS members in YOCs and to measure their physical properties.

\subsubsection{Spatial distribution}

The spatial distribution of cluster members is a classical matter of debate. A recent general review on this topic is contained in S\'anchez \& Alfaro (2009), centred on the distribution of PMS members in YOCs as a primary source of information on the physical properties of the star formation process, both caused from the initial parameters of the parental cloud and from the influence of various environmental conditions. 

The issue is not settled. Here we recall some related observational results. In a study of \object{NGC\,6383}, Fitzgerald et al. (1978) found a ratio of 21 MS to 8 PMS members in the cluster core. Baade (1983) found a weak concentration of PMS members towards cluster center in \object{NGC\,457}, \object{NGC\,7380}, and \object{IC\,1805}, together with the presence of extended halos, which could otherwise  be just weak members, not necessarily PMS stars. In a study of \object{IC\,348}, Lada \& Lada (1995) found a radial decay of both star density and subclustering, without specification of differences between MS and PMS stars. In this same region, Herbig (1998) found a tendency of Weak-lined TTauri (WTT) stars to be more concentrated than Classical TTauri (CTT). On the other hand, he proposed a mean age of 1.4 Myr for the members inside the central 4 arcminutes, where the outer population would have a mean age of 2.8 Myr. He suggested in this case that a young cluster was being observed in projection onto an older background population of PMS stars. Baume et al. (1999), regarding results on \object{NGC\,6231}, proposed the existence of primordial mass segregation, in the sense that lower mass stars would be located at outer locations. In their study of \object{NGC\,1893}, Marco \& Negueruela (2002) found a distinct spatial distribution for MS and PMS members, not related to differences in concentration. They found a relative lack of PMS members at places where the MS members are more numerous. On the other hand, in a study of the Carina OB association, Sartori et al. (2003) found no differences between Massive MS members and PMS members, neither in the spatial distribution, nor in kinematic properties or the age distribution. 

The suggestions of sequential or induced star formation in star forming regions is a topic in itself. Here we mention the results of Prisinzano et al. (2005) on \object{NGC\,~6530}, where they indeed found what was searched for and claimed as not present in the Carina region mentioned before: signs of spatially sequential formation, as a phenomenon different from primordial spatial segregation. The issue of spatially sequential formation is currently the subject of relatively intense study, as has been suggested in several studies of stars-forming regions (Puga et al. 2009, Delgado et al. 2010, and their references).

   \begin{figure}
   \centering
   \includegraphics[angle=0,width=9cm]{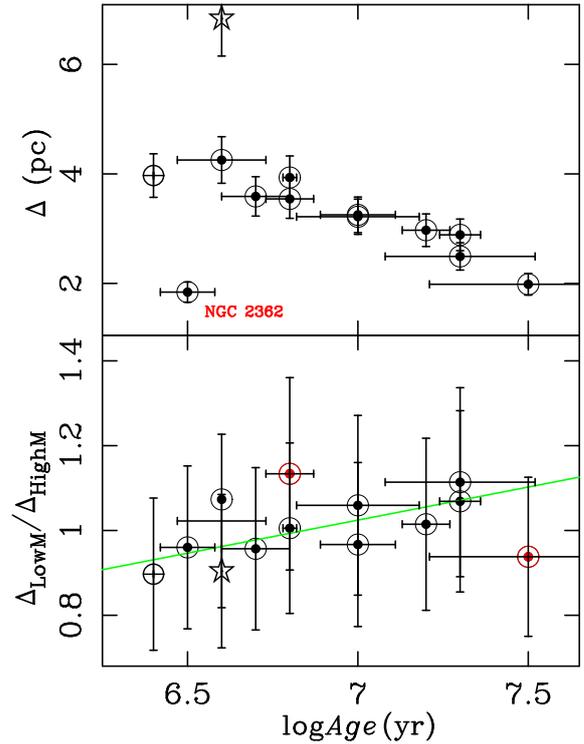}
   \caption{Characteristic clustering scales, $\Delta$ (see text), are plotted versus cluster Age. The upper panel shows $\Delta$ for PMS members of mass above the approximate completeness limit defined in Sect.3.1.2. The relatively more concentrated \object{NGC\,2362} is marked in red. In the lower panel, the ratio between the $\Delta$ values of the lower mass (1-2 $M_\odot$) and higher mass (2-3.5 $M_\odot$) PMS members is shown. Representative points for \object{NGC\,2264} and \object{NGC\,1893} are shown for comparison (crossed circle and star, respectively. See Sect. 2.3). Vertical error bars are calculated with the assumption that the member numbers have absolute error equal to their square root. A linear squares fit with exclusion of \object{NGC\,3293} and \object{Trumpler\,18} (red symbols), and a correlation coefficient of 0.7, is plotted as a straight green line}

              \label{HmLm}
    \end{figure}

So, how do our PMS samples reflect the properties of the spatial structure of the different clusters? The scale of spatial concentration defined in Sect. 3.2 above has been calculated for PMS stars in two separated mass ranges,1-2 $M_\odot$ and 2-3.5 $M_\odot$. These mass intervals were chosen to enhance the possible differences between spatial distributions, and were also limited to the mass range in which we are more confident, avoiding the completeness limit at the lower limit, and the merging at the upper mass limit with possible MS members. 

In the upper panel of Fig.~\ref{HmLm} we plot $\Delta$ versus age for all PMS members of mass above the approximate completeness limit. Symbols for \object{NGC\,2264} and \object{NGC\,1893} are included for comparison, as in previous figures. \object{NGC\,2362} shows a relatively higher concentration for its age. A clear decreasing trend of concentration with age is observed, which can however be affected by some bias due to the distances of the different clusters. In the lower panel of Fig.~\ref{HmLm} we plot the ratio between the concentrations of lower and higher mass PMS members (1-2 $M_\odot$ and 2-3.5 $M_\odot$). In spite of the large error bars, we conjecture a widening of the lower mass PMS members distribution relative to the one for PMS members of higher mass, which amounts to around 20\% in the age range covered. This suggests the presence of dynamical mass segregation, in the sense of a wider distribution for lower mass stars, as the cluster age increases.
 
A formal relation is obtained through a linear least squares fit, also plotted in Fig.~\ref{HmLm}, with a correlation coefficient equal to 0.7. In this fit we exclude the two most deviating points, marked in the plot, which correspond to the clusters \object{NGC\,3293} and \object{Trumpler\,18}. The case of \object{Trumpler\,18} has been commented in connection with the plot in Fig.~\ref{AdvsC}. The field of the cluster is probably containing two superimposed associations, a fact which might influence the measured properties. As commented above, \object{NGC\,3293} has a relatively larger number of massive MS members, as compared to the other clusters in the sample. Some particular features of the star formation process in the cluster could be showing up in its spatial structure. 

\section{Conclusions}

We have presented the results obtained from our procedure in establishing PMS membership in YOCs and simultaneously in measuring the physical parameters of the selected member stars. The procedure is ultimately aimed at checking the performances of the evolutionary  models used in this task, once the third ingredient of spectroscopic observations is added to photometry and models in order to provide an independent confirmation of the membership assignments.

At present, the status of the procedure already appears to be not only promising but actually capable of providing valuable results. This discussion has lead us to suggest some relations between the parameters measured, which agree with previous findings. Independently of these relations, affected by different sources of uncertainty, the main result consists in the presence of regularities which could not plausibly be expected if the selections of members were not well founded. In the present paper we refer to 11 clusters in an age range between 4 and 30 Myr, as determined from comparison of the upper main sequence to  isochrones in the CM diagrams.

   \begin{enumerate}

      \item Two different results of mass values for PMS members in \object{NGC\,2264}, both obtained from the same published photometry with different procedures, suggest that the measurement of masses from comparison to isochrones in the CM diagrams leads to values in better agreement with dynamical masses than those obtained from comparison to isochrones in the theoretical HR diagram.

      \item The derived mass functions for the PMS selected member sequences show in general slopes in agreement with the Salpeter value when using the S00 models, while steeper slopes are obtained when using P99 and Y01 models. A slight increase of the slope with increasing cluster age is suggested. The slopes found for \object{NGC\,3293} and \object{NGC\,2362} are flatter than the Salpeter value. In the case of \object{NGC\,2362}, for which independent membership assignments are available, this finding coincides with other estimates of the mass function. 

       \item The flatter slopes found for the mass functions of \object{NGC\,2362} and \object{NGC\,3293} are reflected in a particularly low ratio between the number of PMS members and the total members number. Whereas \object{NGC\,3293} contains a larger number of massive MS members, as compared to the trend shown by the other clusters, for \object{NGC\,2362} this lower ratio probably  originates in a deficit of lower mass PMS members.

      \item The obtaining of PMS ages with the models used reproduces findings in previous studies, in the sense of increasingly predicted ages from models in the order D97, P99, S00. The ages from the Y01 model are indistinguishable from those in the S00 models, whereas the D97 models predict distinctly lower ages, which amount to several Myr less than the other models. 

      \item The standard deviation of the mean PMS age is in the range 2 to 3 Myr in all clusters. This agrees with general findings on age spread values claimed in several studies of YOCs, but it really needs the assessment of the membership assignments to be confirmed. However, the plot of this quantity versus cluster size, measured as the average of distances between all pairs of PMS members, shows an interesting agreement with previously found scaling relations between age spread and size for star forming regions across a wide range of scales. 

       \item The ratio between characteristic clustering scales for PMS members in the mass ranges 1-2 $M_\odot$ and 2-3.5~$M_\odot$ shows a marginal positive correlation with cluster age. This would suggest the presence of increasing dynamical mass segregation with increasing cluster age, in the sense that lower mass PMS members show a relatively wider distribution for older clusters. Although marginal, we consider this as additional support for the procedure of PMS members selection.

       \item To sum up, CCD wide band photometry can be obtained for practically all young clusters known in the Galactic disk with the use of small to medium size telescopes. For most of them, precise photometry can be obtained for expected PMS cluster members down to spectral type K0, and even fainter, depending on distance and absorption. Is a consistent wealth of information on PMS evolution and physics achievable with these observations? We think that the results in this study show that indeed there is. With our procedure, we simultaneously assign cluster membership from multiband cluster photometry, and measure the mass and age of the individual member candidates. Spectroscopic observations, as a third foot of the tripod in addition to photometry and models, are necessary to achieve a stable set of results. However, the joint use of photometry and models discussed here is already shown to be a promising procedure for the investigation of PMS stars in YOCs.

   \end{enumerate}

\begin{acknowledgements}
This work has been supported by the Spanish Ministerio de Educaci\'on y Ciencia, through grant AYA2007-64052, and by the Consejería de Educación y Ciencia de la Junta de Andalucía, through TIC101. We made use of the NASA ADS Abstract Service and of the WEBDA data base, developed by Jean-Claude Mermilliod at the Laboratory of Astrophysics of the EPFL (Switzerland), and further developed and maintained by Ernst Paunzen at the Institute of Astronomy of the University of Vienna (Austria). This publication makes use of data products from the Two Micron All Sky Survey, which is a joint project of the University of Massachusetts and the Infrared Processing and Analysis Center/California Institute of Technology, funded by the National Aeronautics and Space Administration and the National Science Foundation.
\end{acknowledgements}

\end{document}